\documentclass[journal]{IEEEtran}
\pdfoutput=1
\usepackage[T1]{fontenc}
\usepackage[latin9]{inputenc}
\usepackage{epsfig,psfrag}
\usepackage{graphicx}
\usepackage{amsmath,amsfonts,amssymb,amsxtra,bm,amsthm}
\usepackage{color}
\usepackage{paralist, tabularx}
\usepackage{subfigure}
\usepackage{balance}
\usepackage[adjust]{cite}
\usepackage{enumitem}
\usepackage{xfrac}
\usepackage[disable]{todonotes}  % [disable]
\usepackage{multirow}
\usepackage[final]{changes}
\usepackage{pifont}  % for encircled numbers
\usepackage{subfigure}
\usepackage{siunitx}
\usepackage[hidelinks]{hyperref}
\usepackage{multicol}
\usepackage{tabu}  % allows for thick lines in the table
\usepackage[noend]{algorithmic}
\usepackage{algorithm}

\usepackage{pgfplots, filecontents}
\usetikzlibrary{shapes}
\usepgfplotslibrary{patchplots}

\newcommand{\ve}[1]{\mathbf{#1}}

\newcommand{\rmv}{\hspace*{-.3mm}}

\newcommand{\fig}[1]{\hyperref[#1]{Fig.~\ref*{#1}}}
\newcommand{\tab}[1]{\hyperref[#1]{Tab.~\ref*{#1}}}
\newcommand{\equ}[1]{\hyperref[#1]{(\ref*{#1})}}
\newcommand{\alg}[1]{\hyperref[#1]{Alg.~\ref*{#1}}}
\newcommand{\app}[1]{\hyperref[#1]{App.~\ref*{#1}}}
\newcommand{\chap}[1]{\hyperref[#1]{Chapter~\ref*{#1}}}
\newcommand{\sect}[1]{\hyperref[#1]{Section~\ref*{#1}}}
\newcommand{\highlight}[1]{{\textcolor{black}{#1}}}
\renewcommand{\baselinestretch}{0.92}

\definecolor{WernerRed}{RGB}{233,71,72}
\definecolor{WernerBlue}{RGB}{87,136,175}
\definecolor{WernerGreen}{RGB}{112,190,109}

\graphicspath{{./Fig/}{../../Matlab/Fig/}}
\DeclareGraphicsExtensions{.pdf}
\begin{document}

% -------------------------------------------------------------------
% Title/Subtitle
% -------------------------------------------------------------------
\title{\LARGE{High-Accuracy and Fault Tolerant Stochastic Inner Product Design}}

% -------------------------------------------------------------------
% Authors
% -------------------------------------------------------------------
\author{Werner~Haselmayr,~\IEEEmembership{Member,~IEEE,}
        Daniel~Wiesinger,
        Michael~Lunglmayr,~\IEEEmembership{Member,~IEEE}

%\thanks{Manuscript received month day, 2018; revised month day, 2018; accepted month day, 2018. \IEEEauthorrefmark{1}Corresponding author. 
%W. Haselmayr and D.~Efrosinin are with the Johannes Kepler University Linz, Austria (email: \mbox{werner.haselmayr@jku.at}; \mbox{dmitry.efrosinin@jku.at}). 
%W. Guo is with the School of Engineering, University of Warwick, United Kingdom (email: \mbox{weisi.guo@warwick.ac.uk}).
%}
}

\maketitle

% -------------------------------------------------------------------
% Abstract
% -------------------------------------------------------------------
% ===================================================================
% File: Abstract.tex
% Desc.: 
% Created: 24.01.2018, wh
% Changed: 
% ===================================================================

\begin{abstract}
  In this work, we present a novel inner product design for stochastic computing. Stochastic computing is an emerging computing technique, that encodes a number in the probability of observing a one in a random bit stream. 
  This leads to reduced hardware costs and high error tolerance. The proposed inner product design is based on a two-line bipolar encoding format and applies sequential processing of the input in a central accumulation unit. 
  Sequential processing significantly increases the computation accuracy, since it allows for preliminary cancelation of carry bits. Moreover, the central accumulation unit gives a much better scalability compared to conventional adder tree approaches. We show that the proposed inner product design outperforms a state-of-the-art design in terms of hardware costs for high accuracy requirements and fault tolerance.
  
  %achieves a much better hardware performance for high accuracy requirements and large input vectors compared to state-of-the-art inner product designs.
    %This reduces the hardware costs and increases the scalability compared to conventional adder tree approaches. Moreover, sequential processing enables preliminary cancelation of carry bits, which increases the accuracy.
\end{abstract}


% -------------------------------------------------------------------
% Introduction
% -------------------------------------------------------------------
\section{Introduction}
\label{sec:intro}
% ===================================================================
% File: Introduction.tex
% Desc.: 
% Created: 24.01.2018, wh
% Changed: 
% ===================================================================

\IEEEPARstart{S}{tochastic} computing is an emerging computing technique that encodes a real-valued number into a random bit stream~\cite{Gaines_69}, representing the number as the probability of observing the bit one. This representation allows for a low-complexity implementation of basic arithmetic operations, using only a few logic gates. For instance, the complex multiplier used in conventional binary computing can be replaced by an~AND gate in stochastic computing. Moreover, compared to the binary radix representation, the stochastic representation has a high degree of error tolerance~\cite{Alaghi_17}.

Stochastic computing has been successfully applied in many areas, including decoding of error detection codes, control systems, image processing, filter design and neural networks~(e.g.,~\cite{Alaghi_17,Alaghi_13} and references therein). In many of these applications the inner product is a main building block and, thus, an implementation with low hardware effort and high accuracy is desired. In particular, in neural networks inner products are used to model 
the operation of the neurons~\cite{Lee_17}. Moreover, the FIR filter operation~\cite{Chang_13} and the~DFT/FFT computation~\cite{Yuan_16} are based on the inner product of two vectors.

% Approximate Non-Scaled Adder: Multiplexer-based scaled-adder are replaced by approximate non-scaled adder, which discard all carry weights. This approach offers a highly efficient implementation in terms of HW effort (approx. non-scaled adder can be implemented very efficient), but suffers in accuracy and limits the range of the input vector.
A straightforward implementation of the inner product using an adder tree with multiplexer-based scaled adders, scales down the result, causing severe accuracy loss especially for large vectors. \highlight{To overcome the scaling problem, stochastic to binary conversion was applied in \cite{Ting_14,Kim_16}. However, this comes with the drawback of such a conversion: Additional hardware costs and lower fault tolerance due to the intermediate binary representation. Recently, two approaches have been proposed, addressing the scaling issue within  the stochastic domain~\mbox{\cite{Chang_13,Yuan_16}}}. In~\cite{Chang_13}, an adder tree implementation with multiplexer-based adders using uneven-weights is presented. This reduces the downscaling factor or even scales up the result for certain input values. Unfortunately, the computation of the weights is very complex, and, thus they are often pre-calculated, requiring at least one constant input vector. Moreover, for large vectors the accuracy of the result still degrades due to the growing scaling factor. In~\cite{Yuan_16}, the scaled-adders are replaced by counter-based non-scaled adders using the two-line signed magnitude format~\cite{Toral_00}. It was shown that when applied to a DFT/FFT implementation it achieves a significantly higher accuracy than the approach in~\cite{Chang_13}. Although, the approach in~\cite{Yuan_16} seems very promising there are still some shortcomings. The hardware effort is significantly higher compared to~\cite{Chang_13}, since the non-scaled adder requires more logic gates than a simple scaled adder. Moreover, the accuracy of non-scaled adders is based on the preservation of carry bits, which can be improved by increasing the counter length. Hence, to prevent overflow errors in an adder tree all counter lengths must be increased, leading to a poor scalability in terms of hardware effort. 
%Finally, the implementation in~\cite{Yuan_16}  does not exploit the fact that for some input patterns no carry bits must be generated, which may be lost due to counter overflow.

% Our approach is more robust (TLB format), scalable (central accumulation stage), accuracy (carry canceling)
In this paper, we present a novel stochastic inner product design. We employ the two-line bipolar format \cite{Gaines_69}, enabling a simpler design and achieving higher accuracy compared to~\cite{Yuan_16}. However, we propose simple conversion circuits between the two-line bipolar and the signed magnitude format used in~\cite{Yuan_16}, making the proposed implementation also applicable for the signed magnitude format. 
Instead of an adder tree we use sequential processing of the input in a central accumulation unit, which is realized by a shift-register-based non-scaled adder. The use of a central accumulation unit significantly increases the scalability compared to~\cite{Yuan_16}. Moreover, sequential processing together with the two-line bipolar representation allows for preliminary cancelation of carry bits. This approach reduces the probability of an overflow in the carry register and gives high-accuracy results.

%The carry canceling technique remarkably reduces the number of carry bits that need to be stored in the carry shift registers remarkably. 

%\textcolor{red}{Should we also consider the interface (e.g., SNG) when synthesizing the inner product or just the core datapath (only inner product calculation)}

% -------------------------------------------------------------------
% Stochastic Computing Formats
% -------------------------------------------------------------------
\section{Stochastic Computing Formats}
\label{sec:sc_format}
% ===================================================================
% File: SCFormats.tex
% Desc.: Describe the different formats used in SC and motivates the usage 
%        of the Two Line Bipolar (TLB) format
% Created: 23.05.2018, wh
% Changed: 
% ===================================================================

In this section, we provide an overview on single- and two-line encoding formats used in
stochastic computing. Single-line formats encode a desired number in a single stochastic stream, while two-line formats use two streams.

% In contrast to the single-line encoding formats, two-line representations employ 
%two stochastic streams to encode a desired number.

% ===================================================================
\subsection{Single-Line Encoding Formats}

%In single-line formats, the a desired number is represented using one 
%stochastic bit stream.

% -------------------------------------------------------------------
\subsubsection*{Unipolar Format}
\label{subsec:unip_form}

In the unipolar format, the value of a deterministic number $x \in [0,1]$ is encoded 
in a stochastic bit stream ${X}$ of length $L$, with \cite{Gaines_69}
\begin{align}
  x = \frac{1}{L} \sum\limits_{l=1}^{L} X[l],
\end{align}

\noindent where $X[l] \in \{0,1\}$ denotes the $l$th bit of the bit stream~${X}$.
The precision (representation resolution) of the unipolar format is given by $1/L$. 
Based on this format, basic arithmetic operations can be implemented using simple logic gates~(e.g.,~AND gate for multiplication) \cite{Gaines_69}.%(see Fig.~\ref{fig:basic_sc_circuits}(a)~and~(b)).

% -------------------------------------------------------------------
\subsubsection*{Bipolar Format}
\label{subsec:bip_form}
In contrast to the unipolar format, the bipolar format can also represent negative numbers. This is accomplished through a different interpretation of the stochastic stream.
In this case a number $x \in [-1,1]$ can be represented by a bit stream~${X}$ of length $L$ by \cite{Gaines_69}
\begin{align}
  x & = \frac{1}{L} \sum\limits_{l=1}^{L} 2X[l] - 1.
  \label{eq:bip_conv_repr}
\end{align}

The precision of the bipolar format is given by $2/L$, i.e. half the resolution of the unipolar format. Similar to the unipolar format, the circuits for basic arithmetic operations are very simple \cite{Gaines_69}.%(see Fig. \ref{fig:basic_sc_circuits}(b) and (c)).

%% ---------------------------
%\begin{figure}[t!]
%\centering
  %\includegraphics[scale=0.6]{sc_basic_circuits}
  %\caption{Stochastic circuits for a) a unipolar multiplier, b) a unipolar/bipolar scaled adder and c) a bipolar multiplication.}
  %\label{fig:basic_sc_circuits}
%\end{figure}
%% ---------------------------

% ===================================================================
\subsection{Two-Line Encoding Formats}

%In contrast to the single-line encoding formats, two-line representations employ 
%two stochastic streams to encode a desired number.

% -------------------------------------------------------------------
\subsubsection*{Signed Magnitude Format}
\label{subsec:sgn_mag_form}

In the signed magnitude (SM) format, the sign and magnitude information of a number~\mbox{$x \in [-1,1]$} is carried by the bit streams ${X}_\text{s}$ and ${X}_\text{m}$, respectively. Hence, $x$ can be represented as \cite{Toral_00}
\begin{align}
  x & = \frac{1}{L} \sum\limits_{l=1}^{L} \left(1-2X_\text{s}[l]\right) X_\text{m}[l] ,
  \label{eq:sgn_magn_conv_repr}
\end{align}

\noindent where $X_\text{m}[l]$ and $X_\text{s}[l]$ denote the $l$th bit of the bit streams~${X}_\text{s}$ and ${X}_\text{m}$, respectively.
The SM representation achieves the same resolution as the unipolar format, i.e. $1/L$, while maintaining the same range as the bipolar format. Although, the hardware effort for basic arithmetic operations is higher compared to the unipolar and bipolar format \cite{Toral_00}, it enables an efficient implementation of a non-scaled adder \cite{Yuan_16}. 
Non-scaled adders\footnote{To the best of our knowledge, so far no non-scaled adder has been proposed for the bipolar format.} are very important if multiple successive additions are required (e.g. in an adder tree) since it avoids downscaling. Thus, non-scaled adders are crucial building blocks for an inner product design.
%\textcolor{red}{Non-Scaled adder are important when cascading multiple adder stages due to scaling. Non-scaled adder are not possible with the bipolar format.}

% -------------------------------------------------------------------
\subsubsection*{Two-Line Bipolar Format}
\label{subsec:tlb_form}
% New interpretation makes the format more robust against bit flips
The two-line bipolar (TLB) format uses a different interpretation of the bit streams compared to the SM format. In particular, a number $x\in [-1,1]$ is interpreted as the difference between the numbers $x_\text{p}\in [0,1]$ and $x_\text{n}\in [0,1]$, which are encoded as unipolar bit streams ${X}_\text{p}$ and~${X}_\text{n}$. Hence,~$x$ can be represented as~\cite{Gaines_69}
\begin{align}
  x & = \frac{1}{L}\sum\limits_{l=1}^{L} X_\text{p}[l] - X_\text{n}[l],
  \label{eq:tlb_conv_repr}
\end{align}

\noindent where~$X_\text{p}[l]$ and $X_\text{n}[l]$ denote the $l$th bit of the bit stream~${X}_\text{p}$ and ${X}_\text{n}$, respectively.
The resolution of the TLB format is given by $1/L$. Similar to the SM format, the circuits for the basic arithmetic operations are slightly more complex than for the unipolar and bipolar format \cite{Gaines_69}, but the TLB format also enables an efficient non-scaled adder implementation~(see~Fig.~\ref{fig:tlb_non_scal_add}). As discussed above, non-scaled adders are an important building block for an efficient inner product implementation.

 % However, the format allows an efficient non-scaled adder implementation which is crucial for the inner product implementation proposed in Sec.~\ref{}.

% ---------------------------
\renewcommand{\arraystretch}{1.2}
\begin{table}[t!]
\begin{center}
\caption{TLB and SM Format Conversion}
\label{tab:sm_tlb_conversion}
  \begin{tabular}{c|cc|cc}
    \hline
    %& \multicolumn{2}{c|}{SM Format $W^\text{SM}[i]$} & \multicolumn{2}{c}{$W^\text{TLB}[i] = X_\text{p}[i] - X_\text{n}[i]$} \\
    $X^\text{SM/TLB}[l]$ & $X_\text{s}[l]$ & $X_\text{m}[l]$ & $X_\text{p}[l]$ & $X_\text{n}[l]$ \\ 
    \hline
    $-1$ & 1   & 1 & 0   & 1 \\ 
    $1$  & 0   & 1 & 1   & 0 \\ 
    $0$  & 0/1 & 0 & 0/1 & 0/1 \\ 
    \hline
  \end{tabular}
\end{center}

\end{table}
\renewcommand{\arraystretch}{1.1}
% ---------------------------

% -------------------------------------------------------------------
\subsubsection*{Format Conversion}
\begin{figure}[t!]
\centering
  \includegraphics[scale=0.6]{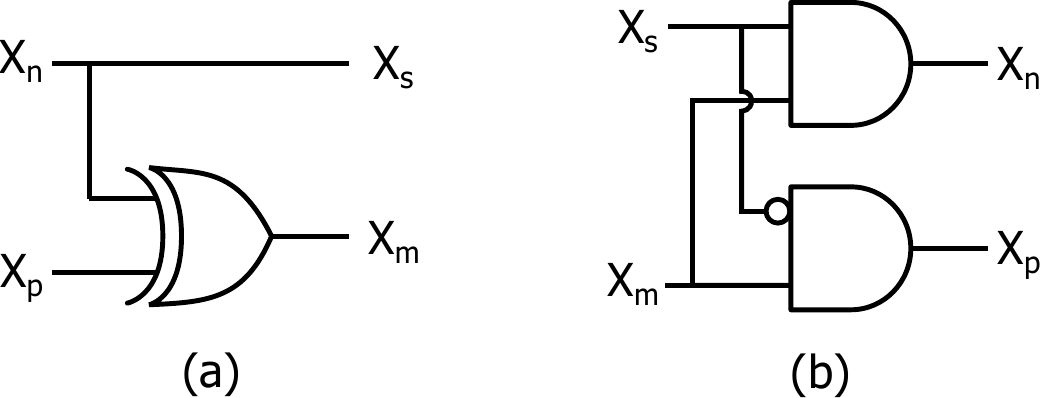}
  \caption{Conversion circuits: (a) TLB to SM; (b) SM to TLB.}
  \label{fig:circuits_sm_tlb_conversion}
\end{figure}

For the conversion between the SM and TLB format we define (see \eqref{eq:sgn_magn_conv_repr})
\begin{align}
  X^\text{SM}[l] = \left(1-2X_\text{s}[l]\right) X_\text{m}[l],
  \label{eq:conv_sm}
\end{align} 

\noindent and (see \eqref{eq:tlb_conv_repr})
\begin{align}
  X^\text{TLB}[l] = X_\text{p}[l] - X_\text{n}[l],
  \label{eq:conv_tlb}
\end{align}

\noindent with $x = 1/L\sum_{l=1}^{L} X^\text{SM/TLB}[l]$. The pairs 
$(X_\text{s}[l],X_\text{m}[l])$ and $(X_\text{p}[l],X_\text{n}[l])$ jointly contribute to $X^\text{SM}[l]$ and $X^\text{TLB}[l]$, respectively, and the elements of 
$X^\text{SM}[l]$ and $X^\text{TLB}[l]$ can only be within the set $\{-1,0,1\}$. Tab.~\ref{tab:sm_tlb_conversion} summarizes these relations, which can be used to derive conversion circuits between the TLB and the SM format as shown in Fig.~\ref{fig:circuits_sm_tlb_conversion}.
For instance, if $X_\text{p}[l] = X_\text{n}[l]$ then $X_\text{m}[l] = 0$ and when
$X_\text{p}[l] \neq X_\text{n}[l]$ then~$X_\text{m}[l] = 1$. Hence, the conversion circuit from the TLB format  to the magnitude stream of the SM format can be realized through an XOR gate~(see~Fig.~\ref{fig:circuits_sm_tlb_conversion}(a)).

% ---------------------------
% ---------------------------

% -------------------------------------------------------------------
% TLB Building Blocks
% -------------------------------------------------------------------
\section{TLB Building Blocks}
\label{sec:tlb_building_blocks}
% ===================================================================
% File: TLBBuildingBlocks.tex
% Desc.: Describes the main building blocks of the TLB format, which are
%        later used for the stochastic inner product
% Created: 23.05.2018, wh
% Changed: 
% ===================================================================

In this section, we present the main building blocks for the TLB format.

% ===================================================================
\subsection{Bit Stream Generator}
\label{subsec:tlb_stream_gen}
As discussed in Sec. \ref{subsec:tlb_form}, the number $x \in[-1,1]$ is defined by the difference between the numbers $x_\text{p}\in[0,1]$ and $x_\text{n}\in[0,1]$, which are encoded into unipolar format bit streams.
Since only the difference matters, $x_\text{p}$ and $x_\text{n}$ are ambiguous and the bit stream generation can be simplified if one of the values is set to zero. Hence, we
represent $x$ as either $x_\text{p}$~\mbox{(if $x \geq 0$)} or $x_\text{n}$~\mbox{(if $x < 0$)}, generate the corresponding stochastic bit stream~${X}_\text{p}$ or~${X}_\text{n}$ and set the other bit stream to zero. 
The conversion of $x_\text{p}$ to ${X}_\text{p}$ or $x_\text{n}$ to ${X}_\text{n}$ can realized using a random generator and a comparator~\cite{Gaines_69}.
%with a randomizer unit~\cite{Gaines_69}, including a random number generator\footnote{Typically, a linear feedback shift register (LFSR) is used as random generator.} and a comparator.

% ===================================================================
\subsection{Multiplier}
\label{subsec:tlb_mult}
\begin{figure}[t!]
\centering
  \includegraphics[scale=0.5]{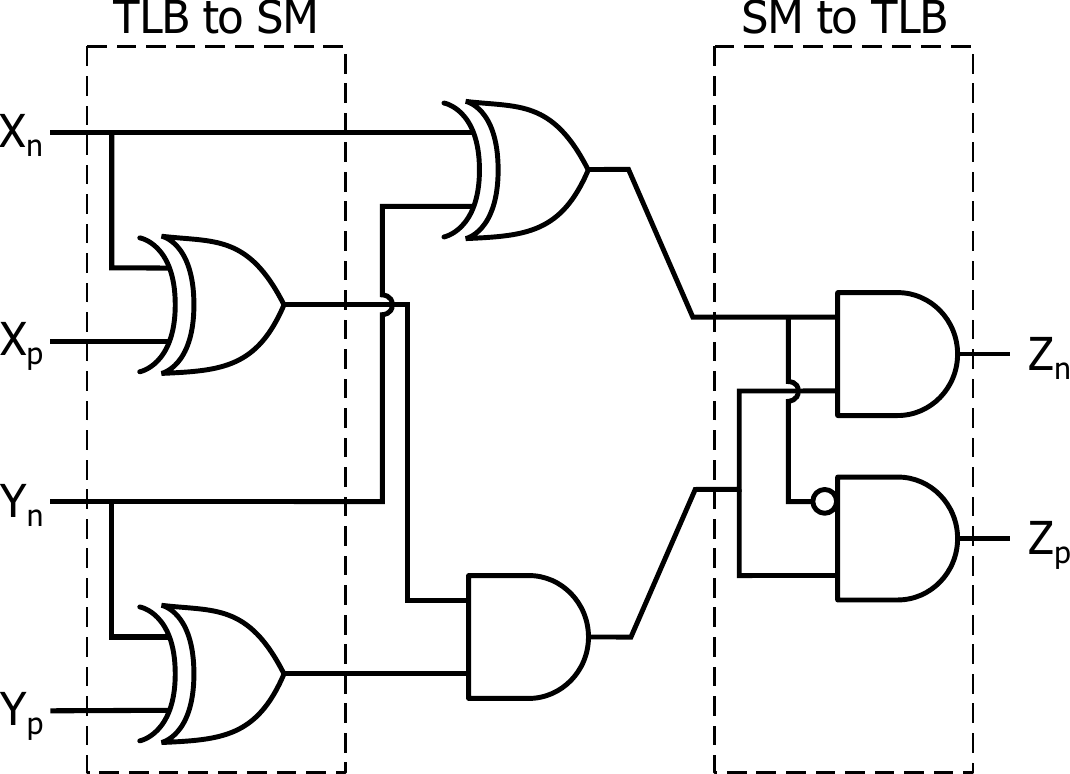}
  \caption{Circuit of the stochastic multiplier for the TLB format.}
  \label{fig:tlb_mult}
\end{figure}

\highlight{A first circuit of a multiplier for the TLB format has been proposed in~\cite{Gaines_69}. In Fig.~\ref{fig:tlb_mult} we present an alternative multiplier circuit.} The core circuit corresponds to the multiplier for the SM format (XOR and AND gate) \cite{Yuan_16} and the interface corresponds to the conversion circuit between TLB and SM format shown in Fig. \ref{fig:circuits_sm_tlb_conversion}. It is important to note that the presented circuit is only used for illustration purpose and a more simple design can be obtained through logic optimization.

% ===================================================================
\subsection{Non-Scaled Adder}
\label{subsec:tlb_non_scal_adder}
\begin{figure}[t!]
\centering
  \includegraphics[scale=0.55]{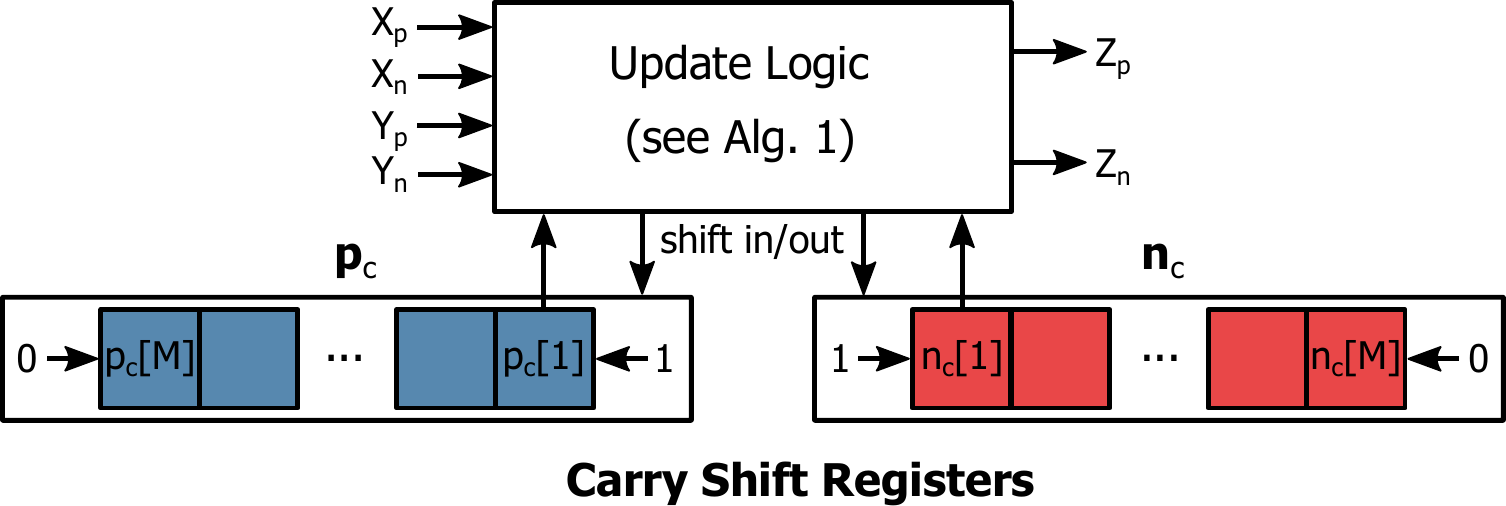}
  \caption{Circuit of the shift-register-based stochastic non-scaled adder for the TLB format.}
  \label{fig:tlb_non_scal_add}
\end{figure}

\highlight{To the best of our knowledge, only scaled adders have been proposed for the TLB format (see~\cite{Gaines_69}). Thus, we present a novel shift-register-based non-scaled adder\footnote{Although also a counter-based non-scaled adder can be used, we propose a shift-register-based implementation because of its higher fault tolerance~\cite{Ting_17}.} as shown in Fig.~\ref{fig:tlb_non_scal_add}. The
circuit consists of an update logic and carry shift registers $\ve{p}_\text{c}$ and $\ve{n}_\text{c}$, each of size $M$.
The update logic must consider many different cases, including the preservation and cancellation of carry bits in the carry shift registers. For example, let's consider the numbers $x$, $y$ and their sum~$z$. According to (6), these number can be represented as streams \mbox{$X[l], Y[l], Z[l] \in \{-1,0,1\}$}. However, since each element of the result $Z[l]$ can only be within the set $\{-1,0,1\}$, $Z[l]$ is not only the sum of $X[l]$ and $Y[l]$, but the effect of carry must be considered. If both $X[l]$ and $Y[l]$ are either $1$ or $-1$, $Z[l]$ is either $1$ or $-1$ and a carry 1 ($\ve{p}_c$ shift in) or $-1$ ($\ve{n}_c$ shift in) should be stored  
in the carry shift registers for the next calculation. However, it is also possible that the current carry bit cancels a stored carry bit from a previous calculation, e.g. a generated carry $1$ cancels a stored carry $-1$ ($n_c[1] =1$). The update logic algorithm given in Alg.~1 takes into account all this different scenarios. Please note that the shift in operation denotes that a one (carry bit) is shifted into the register on one side,
while the shift out operation denotes that a zero is shifted into the register on the other side, i.e. a carry bit is shifted out of the register.}

% ---------------------------
\renewcommand\footnoterule{}     
\begin{algorithm}[] %H
\caption{Update Logic for Non-Scaled Adder}%\protect\footnotemark}
\begin{algorithmic}[1]
\renewcommand{\algorithmicrequire}{\textbf{Input:}}
\renewcommand{\algorithmicensure}{\textbf{Output:}}
\REQUIRE ${X}$, ${Y}$
%\ENSURE $\m{Z}$
 \\ \noindent \textit{Initialization}: $\ve{n}_c = \ve{0}$, $\ve{p}_c = \ve{0}$
  \FOR {$i = 1$ to $L$}
    \IF {$X[l] + Y[l] = 0$}
       \STATE $Z[l] \leftarrow p_c[1] - n_c[1]$; $\ve{p}_c$ and $\ve{n}_c$ shift out 
    \ELSIF{$X[l] + Y[l] = 1$}
      \STATE $Z[l] \leftarrow 1 - n_c[1]$; $\ve{n}_c$ shift out 
    \ELSIF{$X[l] + Y[l] = -1$}
      \STATE $Z[l] \leftarrow p_c[1]- 1$; $\ve{p}_c$ shift out
    \ELSIF{$X[l] + Y[l] = 2$}
      \STATE $Z[l] \leftarrow 1$%
      \IF {$n_\text{c}[1] = 1$}
        \STATE  $\ve{n}_c$ shift out
      \ELSE
        \STATE  $\ve{p}_c$ shift in
      \ENDIF
    \ELSIF{$X[l] + Y[l] = -2$}
      \STATE $Z[l] \leftarrow -1$%; $\ve{n}_c$ shift in
      \IF {$p_\text{c}[1] = 1$}
        \STATE  $\ve{p}_c$ shift out
      \ELSE
        \STATE  $\ve{n}_c$ shift in
      \ENDIF      
  \ENDIF
\ENDFOR
\RETURN ${Z}$ 
\end{algorithmic} 
\label{alg:update_logic_non_scaled_adder}
\end{algorithm}
% ---------------------------
%\footnotetext{Shift in/out: A one/zero is shifted into a particular carry shift register~\mbox{(see~Fig.~\ref{fig:tlb_non_scal_add})}.}

% -----------------------------------------------
% Recursive equations for "Cancelor"
%The recursive relation (\textit{logistic map}) can be written in explicit form as
%\begin{align}
 %p_n = \ve{e}\m{T}^{\lceil n \rceil} \ve{p}^\text{T}, \quad n = 1,\ldots, M
%\end{align}
%
%with $\ve{e} = [1,0,0, \ldots, 0]^\text{T}$ and $\ve{p} = [p^1, p^2, \ldots, p^{2^{n/2}}]^\text{T}$, with $p = p_0 = 0.25$. Moreover, the 
%matrix $\m{T}$ is defined by 
%\begin{align}
  %T_{l,k}=\begin{cases}
    %(-1)^{k-l}\binom{l}{k-l} , & \forall l,k \in \mathbb{N}: l \geq 1, l \leq k \leq 2l.\\
    %0, & \text{otherwise}.
  %\end{cases}
%\end{align}
% -----------------------------------------------

% ---------------------------
% ---------------------------

% -------------------------------------------------------------------
% Stochastic Inner Product
% -------------------------------------------------------------------
\section{Stochastic Inner Product Desgin}
\label{sec:stoch_inn_prod}
% ===================================================================
% File: StochasticInnerProduct.tex
% Desc.: Describes the implementation of the stochastic inner product
% Created: 23.05.2018, wh
% Changed: 
% ===================================================================

\begin{figure*}[t!]
  %\centering
  \includegraphics[width=2\columnwidth]{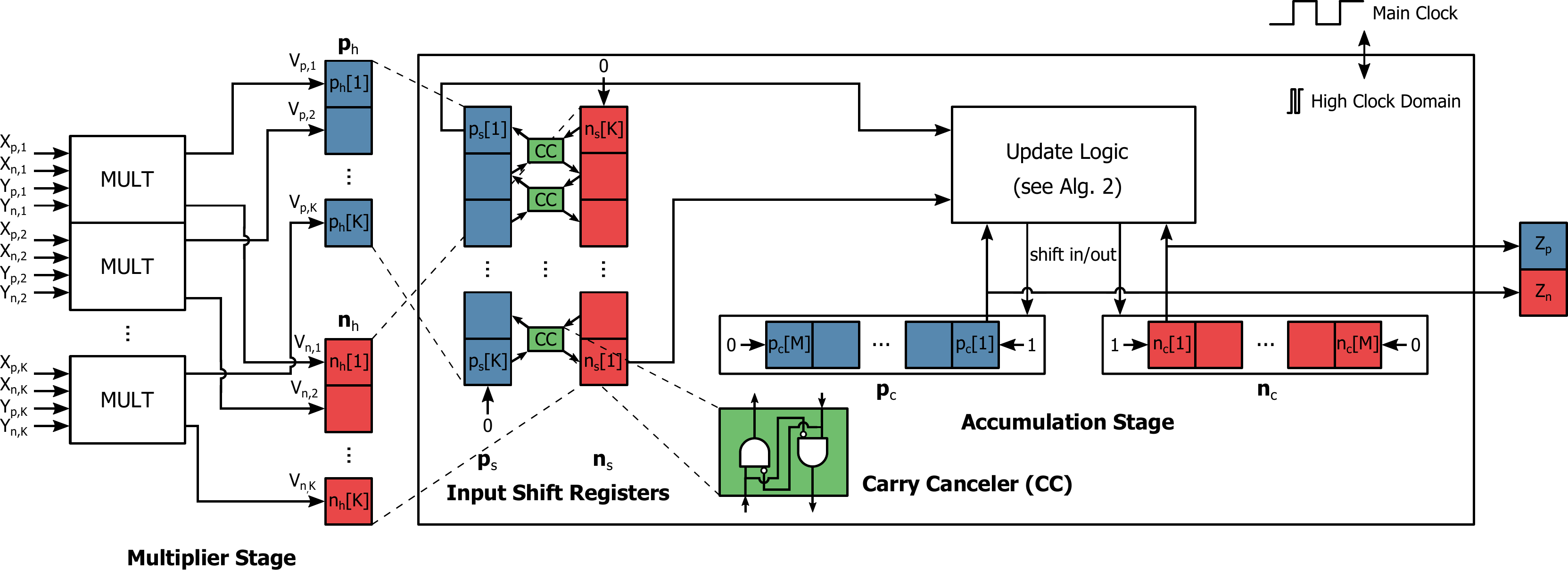}
  \caption{Architecture of the novel stochastic inner product desgin.}
  \label{fig:direct_scalar_product_overview}
\end{figure*}

In this section, we present the stochastic inner product implementation. The architecture is shown in Fig. \ref{fig:direct_scalar_product_overview}, including a multiplier stage, input shift registers with carry canceling and an accumulation stage. For the following description we consider the computation of the inner product between the vectors $\ve{x} = [x_1,\ldots, x_K]^\text{T}$ and $\ve{y} = [y_1, \ldots, y_K]^\text{T}$ given by
\begin{align}
  z = \langle \ve{x}, \ve{y} \rangle =\ve{x}^\text{T}\ve{y} = \sum\limits_{k=1}^{K} x_k y_k,
\end{align}

\noindent with $x_k, y_k \in[-1,1]$. The numbers $x_k$ and $y_k$ are encoded in the stochastic bit streams $(X_{\text{p},k},X_{\text{n},k})$ and $(Y_{\text{p},k},Y_{\text{n},k})$ using the TLB format.

% ===================================================================
\subsubsection{Multiplier Stage} This stage performs the multiplication of 
the individual entries of the input vectors, i.e. $v_k = x_k y_k$, using $K$
stochastic multipliers as shown in Fig.~\ref{fig:tlb_mult}. Each multiplier has the streams $({X}_{\text{p},k},{X}_{\text{n},k})$ and $({Y}_{\text{p},k},{Y}_{\text{n},k})$ at its input and generates the streams  $({V}_{\text{p},k}, {V}_{\text{n},k})$. The individual bits of the output streams are stored for one clock cycle of the main clock in the input hold registers $\ve{p}_\text{h}$ and $\ve{n}_\text{h}$, respectively. \highlight{These registers prevent intermediate results from propagating from the main clock domain (multiplier stage) into the higher clock domain (input shift registers, accumulation stage).}
%The synchronization between the main clock and the higher clock domain is accomplished through three flip-flops (see top left in Fig. \ref{fig:direct_scalar_product_overview}), at which the last flip-flop is used for rising edge detection.

% ---------------------------
% ---------------------------

% ---------------------------
% ---------------------------

% ===================================================================
\color{black}
\subsubsection{Input Shift Registers with Carry Canceling} Upon a rising edge of the main clock, the elements of the input hold registers $\ve{p}_\text{h}$ and $\ve{n}_\text{h}$ are copied into the input shift registers $\ve{p}_\text{s}$ and $\ve{n}_\text{s}$, following the mapping: ${p}_{h}[1] \rightarrow {p}_{s}[1]$, ${p}_{h}[2] \rightarrow {p}_{s}[2]$, etc., and 
  ${n}_{h}[1] \rightarrow {n}_{s}[K]$, ${n}_{h}[2] \rightarrow {n}_{s}[K-1]$, etc. This type of mapping increases the probability that ones are canceled by the so-called carry canceler (CC).
  The aim of the CC is to reduce the number of ones that are shifted towards the accumulation stage, which reduces the probability of an overflow of the carry shift registers. Hence, this improves the 
  accuracy of the inner production calculation. The CC circuit is shown in Fig.~\ref{fig:direct_scalar_product_overview}, where the outputs are zero if both inputs are one and otherwise the outputs follows the inputs.
  The diagonal elements of the input shift registers are connected by the CC (see Fig.~\ref{fig:direct_scalar_product_overview}) and, thus, the value of the $k$th register element after the shifting operation is given by
\begin{equation}
\begin{aligned}
  p_\text{s}[k] & = p_\text{s}[k+1]\overline{n_\text{s}}[K-k+1] \\
  n_\text{s}[k] & = n_\text{s}[k+1]\overline{p_\text{s}}[K-k+1],
  \label{eq:bool_fct_cc}
\end{aligned}
\end{equation}

\noindent where $\overline{(\cdot)}$ denotes the negation operator. Please note that \eqref{eq:bool_fct_cc} corresponds to the Boolean function of the CC.
%and  $p_\text{s}[k]$ and $n_\text{s}[k]$ denote the $k$th element of the input shift registers $\ve{p}_\text{s}$ and $\ve{n}_\text{s}$, respectively.
In particular, the canceling procedure is as follows: Upon a rising edge of the higher clock, the CC output is written into the next register element. This corresponds either to shifting the value of the previous element to the next element (normal shift operation) or writing zeros (carry canceling). 

\begin{figure}[t!]
  %%\begin{center}
  \begin{tikzpicture}
  %\begin{semilogyaxis}
  \begin{axis}[compat=newest, 
    width=0.95\columnwidth, height =0.5\columnwidth, grid,
    ylabel={\small{$P_\text{p} = P_\text{n}$}}, 
    xlabel={\small{Input Shift Register Length $K$}}, 
    xmin = 0,
    xmax = 64,
    ymin = 0,
    ymax = 0.3,
    xtick = {0, 20, 40, 60},
    legend pos=south west, 
    legend columns = {1},
    legend cell align=left,
    label style={font=\small},
    tick label style={font=\small},
  ]
  %\addplot[color=black,very thick] table[x index =0, y index =1] {./dat/CancelorPerformance.dat};
  %\addlegendentry{}
  \addplot[color=WernerBlue,very thick,mark=o, mark repeat=3, mark phase=0] table[x index =0, y index =1] {CancelorPerformance1e5.dat}; % y index = 2
  \addlegendentry{\footnotesize{Opposite Direction}}
  \addplot[color=WernerRed,very thick, mark=o, mark repeat=3, mark phase=0] table[x index =0, y index =3] {CancelorPerformance1e5.dat};  % y index = 4
  \addlegendentry{\footnotesize{Same Direction}}
  %\addplot[color=WernerBlue, only marks, very thick, forget plot, mark=o, mark repeat=3, mark phase=0] table[x index =0, y index =1] {./dat/CancelorPerformance1e5.dat};
  %\addplot[color=WernerRed, only marks, forget plot, very thick, mark=o, mark repeat=3, mark phase=0] table[x index =0, y index =3] {./dat/CancelorPerformance1e5.dat};
  %\addplot[color=WernerRed,very thick] table[x index =0, y index =4] {./dat/CancelorPerformance.dat};
  %\addlegendentry{ Novel SMax }
  %\end{semilogyaxis}
  \end{axis}
  \end{tikzpicture}
  \caption{Average probability that ones are shifted towards the accumulation stage  $P_\text{p},\,P_\text{n}$ versus the input shift register length $K$, assuming that the probability 
  that ones are copied from the input hold registers to the input shift registers is 0.5.}
  \label{fig:perf_canceler}
  %\end{center}
  \end{figure}
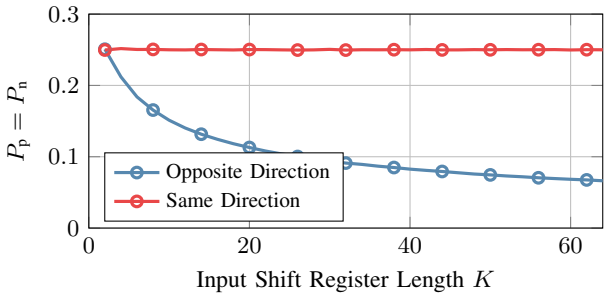

The elements $p_\text{s}[1]$ and $n_\text{s}[1]$ are sequentially shifted to the accumulation stage using a higher clock compared to the main clock. 
Please note that the input shift registers $\ve{p}_\text{s}$ and $\ve{n}_\text{s}$ are shifted in opposite directions (see Fig.~\ref{fig:direct_scalar_product_overview}),  which reduces the probability that ones are shifted to the accumulation stage compared to shifting in the same direction. This is because in case of shifting in the same direction the CC has only an effect after the first shifting operation. We validated the impact of 
the shifting direction through bit-true simulations. Therefore we evaluated the average probability that ones are shifted towards the accumulation stage during sequential processing of the entire input shift registers $\ve{p}_\text{s}$ and $\ve{n}_\text{s}$, which is given by
\begin{align}
  P_x \rmv = \rmv \frac{1}{K}\sum_{j=1}^{K} \Pr(x_\text{s}^{(j)}[1] = 1),
  \label{eq:avg_shif_perf}
\end{align}

\noindent with $x \in\{\text{p},\text{n}\}$ and $\Pr(x_\text{s}^{(j)}[1] = 1)$ denotes the probability that that a one is shifted towards the accumulation stage after the $j$th shift operation. The results are shown in Fig.~\ref{fig:perf_canceler}, confirming
that for  $K \geq 2$ shifting in the opposite direction should be preferred to shifting in the same direction.

\color{black}
% ---------------------------

% ===================================================================
\subsubsection{Accumulation Stage} 
\highlight{The accumulation stage corresponds to a shift-register-based non-scaled adder (see \mbox{Sec.~ \ref{subsec:tlb_non_scal_adder}}), which accumulates the output of the input shift registers in the carry shift registers. Similar to the non-scaled adder, the accumulation stage considers many different scenario, including the preservation and cancellation of carry bits. The corresponding algorithm is given in Alg.~\ref{alg:update_logic_non_scaled_adder_inner_prod}}

%The accumulation stage corresponds to a  shift-register-based stochastic non-scaled adder similar to Sec. \ref{subsec:tlb_non_scal_adder}. It accumulates the results of the multiplication stage. The algorithm for the update logic is given in Alg.~\ref{alg:update_logic_non_scaled_adder_inner_prod}. 
% ---------------------------
\renewcommand\footnoterule{}
\begin{algorithm}[]
\caption{Update Logic for Accumulation Stage}
\begin{algorithmic}[1]
\renewcommand{\algorithmicrequire}{\textbf{Input:}}
\renewcommand{\algorithmicensure}{\textbf{Output:}}
\REQUIRE $X = p_\text{s}[1] - n_\text{s}[1]$, $C = p_\text{c}[1] - n_\text{c}[1]$
%\ENSURE $\m{W}_3$
 %\ \noindent \textit{Initialization}: $\ve{n}_c = \ve{0}$, $\ve{p}_c = \ve{0}$
  \FOR {$k = 1$ to $K$}
    \IF{$X= 0$ \& $C = 0$}
       \STATE $\ve{p}_\text{c}$ and $\ve{n}_\text{c}$ shift out
    \ELSIF{$X = 1$}
      \IF{$C = 0$ }
        \STATE $\ve{p}_\text{c}$ shift in ($p_\text{c}[1] \rmv = \rmv n_\text{c}[1] \rmv = \rmv 0$) or \\ $\ve{n}_\text{c}$ shift out \mbox{($p_\text{c}[1] \rmv = \rmv n_\text{c}[1] \rmv = \rmv 1$)}
      \ELSIF{$C = - 1$}
        %\STATE $\ve{p}_\text{c}$ and $\ve{n}_\text{c}$ shift out
        \STATE $\ve{n}_\text{c}$ shift out
      \ELSE 
        \STATE $\ve{p}_\text{c}$ shift in
      \ENDIF
    \ELSIF{$X = -1 $}
      \IF{$C = 0$ }
        \STATE $\ve{n}_\text{c}$ shift in ($p_\text{c}[1] \rmv = \rmv n_\text{c}[1] \rmv = \rmv 0$) or \\ $\ve{p}_\text{c}$ shift out \mbox{($p_\text{c}[1] \rmv = \rmv n_\text{c}[1] \rmv = \rmv 1$)}
      \ELSIF{$C = 1$}
        \STATE $\ve{p}_\text{c}$ shift out
      \ELSE % $C = 1$
        \STATE $\ve{n}_\text{c}$ shift in
      \ENDIF
  \ENDIF
  \STATE Shift $\ve{p}_s$ and $\ve{n}_s$ %\, $\rightarrow$ \, $X = p_\text{h}[1] - n_\text{h}[1]$
  \STATE $X = p_\text{s}[1] - n_\text{s}[1]$
\ENDFOR
%\RETURN $\m{W}_3$ 
\end{algorithmic} 
\label{alg:update_logic_non_scaled_adder_inner_prod}
\end{algorithm}
% ---------------------------

It is important to note that the sequential processing of the input shift registers must be finished upon the next rising edge of the main clock. Then, the input shift registers are loaded with the next inputs from the input hold registers. Moreover, the entries $p_\text{c}[1]$ and $n_\text{c}[1]$ of the carry shift registers are shifted to the output flip-flops corresponding to the $l$th bit in the output stochastic streams i.e. ($Z_{\text{p}}[l],Z_{\text{n}}[l]$).

%It is important to note that the carry canceler in the input shift register together with 
%the updated logic for the non-scaled adder significantly reduces the number of produced carry bits. Hence, the probability for an overflow of the carry shift register can be reduced, ensuring high calculation accuracy of the inner product.
%
%In the next section, we show that the proposed inner product outperforms state-of-the-art inner product designs, when the accuracy requirements are high and large input input vectors are used.

% -------------------------------------------------------------------
% Performacne Analysis
% -------------------------------------------------------------------
\section{Performance Analysis}
\label{sec:perf_analysis}
% ===================================================================
% File: PerformanceAnalysis.tex
% Desc.: Performance Analysis of the proposed stochastic inner product
% Created: 23.05.2018, wh
% Changed: 
% ===================================================================

% NOTES:
% -Should we include some concrete examples (this procedure is x% better than the other) as it was done in the thesis. 
% - Hardware costs for the interface are neglected!

%The state-of-the-art implementation applies the signed magnitude format and uses an adder tree with counter-based non-scaled adders \cite{Yuan_16}.

In this section, we compare the proposed inner product design with the state-of-the-art design presented in~\cite{Yuan_16} in terms resource utilization and fault tolerance for different accuracy requirements. For the comparison, we only consider the inner product calculation and omit the costs for the stochastic stream generation and the back conversion, since they are similar for both approaches. 

We define the computation accuracy by the root mean square error (RMSE) given by $\text{RMSE} \rmv=\rmv \sqrt{\text{mean}(|\hat{z} - z|)}$, where $z$ denotes the true inner product result (double-precision floating point) and~$\hat{z}$ corresponds to the results of the particular stochastic implementation. The accuracy is controlled by the carry shift register length and the counter length for the novel and the state-of-the-art design, respectively. For all investigations we fixed the length of the stochastic stream to~$L\rmv=\rmv 10^{4}$.

\begin{figure}[t!]
  %%\begin{center}
  \begin{tikzpicture}
  %\begin{semilogyaxis}
  \begin{axis}[compat=newest, 
    width=0.95\columnwidth, height =.5\columnwidth, grid,
    ylabel={\small{No. Registers}}, 
    xlabel={\small{Input Vector Length $K$ }}, 
    xmin = 0,
    xmax = 64,
    ymin = 0,
    ymax = 300,
    xtick = {0, 20, 40, 60},
    ytick = {0, 100, 200, 300},
    label style={font=\small},
    tick label style={font=\small},
    legend pos=north west, 
    legend columns = {1},
    legend cell align=left,
    legend style={row sep=0.05pt},
    legend image post style={black} % make legend line BLACK!!!
  ]
  \addplot[color=WernerBlue,very thick] table[x index =0, y index =1] {InnerProdComplexityRegisters.dat};
  \addlegendentry{\footnotesize{Novel design}}
  \addplot[color=WernerRed,very thick,forget plot] table[x index =0, y index =2] {InnerProdComplexityRegisters.dat};
  %\addlegendentry{\footnotesize{Novel stoch. IP}}
  \addplot[color=WernerGreen,very thick,forget plot] table[x index =0, y index =3] {InnerProdComplexityRegisters.dat};
  %\addlegendentry{\footnotesize{Novel stoch. IP}}
  
  \addplot[color=WernerBlue,very thick,dashed] table[x index =0, y index =4] {InnerProdComplexityRegisters.dat};
  \addlegendentry{\footnotesize{SoA design \cite{Yuan_16}}}
  \addplot[color=WernerRed,very thick,forget plot,dashed] table[x index =0, y index =5] {InnerProdComplexityRegisters.dat}; 
   %\addlegendentry{\footnotesize{SoA stoch. IP}}
  \addplot[color=WernerGreen,very thick,forget plot,dashed] table[x index =0, y index =6] {InnerProdComplexityRegisters.dat}; 
   %\addlegendentry{\footnotesize{SoA stoch. IP}}
  %\end{semilogyaxis}
  \end{axis}
  \end{tikzpicture}
  \caption{Number of registers required for the novel  and state-of-the-art implementation. Blue, red and green curves correspond to $\text{RMSE} \leq 0.1$, $\text{RMSE} \leq 0.05$ and $\text{RMSE} \leq 0.02$.}
  \label{fig:inner_prod_compl_reg}
  %\end{center}
  \end{figure}
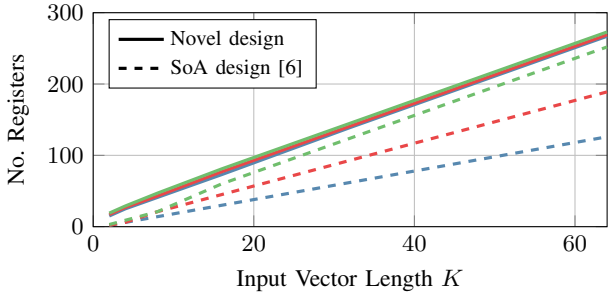

We determined the resource utilization for both implementations through synthesis for an Altera Cyclone IV  EP4CE115 FPGA. Figs. \ref{fig:inner_prod_compl_le} and \ref{fig:inner_prod_compl_reg} show the minimum number of logic elements (combinational logic) and registers that are required to achieve a certain computation accuracy.  We observe from Fig. \ref{fig:inner_prod_compl_le} that the logic element utilization of the proposed design is much better compared to the state-of-the-art design, especially for large input vectors and if high computation accuracy is required. Moreover, we observe from Fig.~\ref{fig:inner_prod_compl_reg} that if low accuracy is sufficient, the state-of-the-art approach outperforms the novel design in terms of register utilization. This is because the approach in~\cite{Yuan_16} requires no hold circuit at the input (input hold registers) or sequential processing storage (input shift registers). Interestingly, for the novel design the logic element and register utilization is almost independent of the accuracy requirements, while it increases for the state-of-the-art implementation. 
This means that for the proposed design the additional hardware effort (larger carry shift registers) to achieve a better accuracy is insignificant.
%That means the accuracy of the proposed design depends only on the length of the stochastic stream.

%This behavior is an unique benefit of the proposed design, since it allows to control the accuracy only by changing the length of the stochastic input streams.

% ---------------------------
  \begin{figure}[t!]
  %%\begin{center}
  \begin{tikzpicture}
  %\begin{semilogyaxis}
  \begin{axis}[compat=newest, 
    width=0.95\columnwidth, height =.5\columnwidth, grid,
    ylabel={\small{No. Logic Elements}}, 
    xlabel={\small{Input Vector Length $K$ }}, 
    xmin = 0,
    xmax = 64,
    ymin = 0,
    ymax = 2500,
    xtick = {0, 20, 40, 60},
    ytick = {0, 500, 1500, 2500},
    label style={font=\small},
    tick label style={font=\small},
    legend pos=north west, 
    legend columns = {1},
    legend cell align=left,
    legend style={row sep=0.05pt},
    legend image post style={black}  % make legend line BLACK!!!
  ]
  \addplot[color=WernerBlue,very thick] table[x index =0, y index =1] {InnerProdComplexityLogicElements.dat};
  \addlegendentry{\footnotesize{Novel design}}
  \addplot[color=WernerRed,very thick,forget plot] table[x index =0, y index =2] {InnerProdComplexityLogicElements.dat};
  %\addlegendentry{\footnotesize{Novel stoch. IP}}
  \addplot[color=WernerGreen,very thick,forget plot] table[x index =0, y index =3] {InnerProdComplexityLogicElements.dat};
  %\addlegendentry{\footnotesize{Novel stoch. IP}}
  
  \addplot[color=WernerBlue,very thick,dashed] table[x index =0, y index =4] {InnerProdComplexityLogicElements.dat};
  \addlegendentry{\footnotesize{SoA design \cite{Yuan_16}}}
  \addplot[color=WernerRed,very thick,forget plot,dashed] table[x index =0, y index =5] {InnerProdComplexityLogicElements.dat}; 
  %\addlegendentry{\footnotesize{SoA stoch. IP}}
  \addplot[color=WernerGreen,very thick,forget plot,dashed] table[x index =0, y index =6] {InnerProdComplexityLogicElements.dat}; 
  %\addlegendentry{\footnotesize{SoA stoch. IP}}

  %\addplot[color=WernerBlue,very thick,mark=square] table[x index =0, y index =7] {./dat/InnerProdComplexityLogicElements.dat};
  %\addlegendentry{\footnotesize{FP IP}}
  %\addplot[color=WernerRed,very thick,mark=square,forget plot] table[x index =0, y index =9] {./dat/InnerProdComplexityLogicElements.dat};
  %\addlegendentry{\footnotesize{FP IP}}  
  %\end{semilogyaxis}
  \end{axis}
  \end{tikzpicture}
  \caption{Number of logic elements required for the novel and state-of-the-art implementation. Blue, red and green curves correspond to $\text{RMSE} \leq 0.1$, $\text{RMSE} \leq 0.05$ and $\text{RMSE} \leq 0.02$.}
  \label{fig:inner_prod_compl_le}
  %\end{center}
  \end{figure}
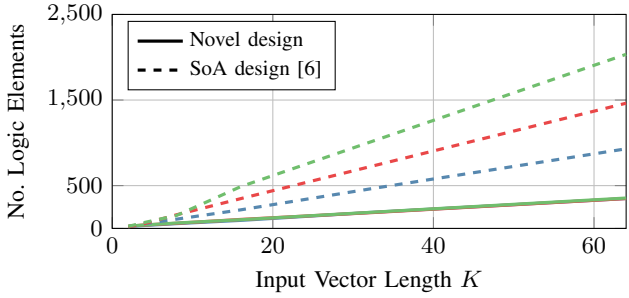
% ---------------------------

% ---------------------------
% ---------------------------

% ---------------------------
  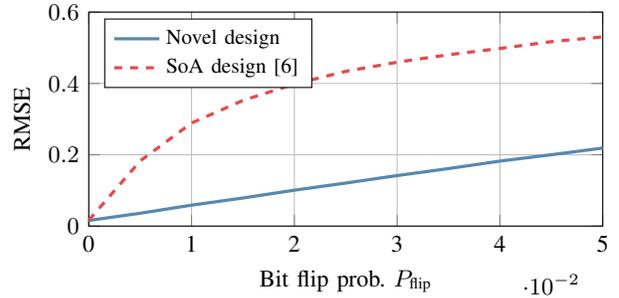
\begin{figure}[t!]
  %%\begin{center}
  \begin{tikzpicture}
  %\begin{semilogyaxis}
  \begin{axis}[compat=newest, 
    width=0.95\columnwidth, height =.5\columnwidth, grid,
    ylabel={\small{RMSE}}, 
    xlabel={\small{Bit flip prob. $P_\text{flip}$}}, 
    xmin = 0,
    xmax = 0.05,
    ymin = 0,
    ymax = 0.6,
    xtick = {0, 0.01, 0.02, 0.03, 0.04, 0.05},
    xticklabels = {0, 1, 2, 3, 4, 5},
    ytick = {0,0.2, 0.4, 0.6, 0.8, 1},
    label style={font=\small},
    tick label style={font=\small},
    legend pos=north west, 
    legend columns = {1},
    legend cell align=left,
    legend style={row sep=0.05pt}
  ]
  \addplot[color=WernerBlue,very thick] table[x index =0, y index =3] {InnerProdRobustness.dat};
  \addlegendentry{\footnotesize{Novel design}}
  \addplot[color=WernerRed,very thick, dashed] table[x index =0, y index =2] {InnerProdRobustness.dat};
  \addlegendentry{\footnotesize{SoA design \cite{Yuan_16}}}
  %\addplot[color=WernerGreen,very thick,mark=square] table[x index =0, y index =3] {./dat/InnerProdRobustness.dat};
  %\addlegendentry{\footnotesize{FP IP}}
  %\end{semilogyaxis}
  \end{axis}
  \end{tikzpicture}
  \caption{Robustness against bit flips of the novel and state-of-the-art inner product design.}
  \label{fig:inner_prod_robust}
  %\end{center}
  \end{figure}
% ---------------------------

Fig.~\ref{fig:inner_prod_robust} compares the fault tolerance of the novel and state-of-the-art inner product design. Therefore, we randomly flipped a bit in the carry shift registers or the counters in the adder tree with probability $P_\text{flip}$. 
This approach gives a good approximation of the fault tolerance for the entire design, 
since failures in the storage can also be interpreted as bit flips coming from the combinational logic. We used  input vectors of length \mbox{$K\rmv=\rmv 16$} and started with the computation accuracy \mbox{$\text{RMSE}\rmv=\rmv 0.02$}. This requires a carry shift register length of $6$ and a counter length of $4$. We observe that the proposed design is much more robust against bit flips than the state-of-the-art implementation. This is mainly because we use a shift-register-based approach, rather than a counter-based approach. 

For the novel design it is important to note that although the high clock domain can operate nearly at maximum platform speed (short critical path), the main clock is reduced by the input vector length $K$. However, this issue can be easily solved through parallelization of the sequential processing step, using multiple inner product cores.

% -------------------------------------------------------------------
% Conclusions
% -------------------------------------------------------------------
\section{Conclusions}
\label{sec:concl}
% ===================================================================
% File: Conclusionsc.tex
% Desc.: 
% Created: 24.01.2018, wh
% Changed: 
% ===================================================================

In this work, we proposed a novel stochastic inner product design. In contrast to state-of-the art adder tree implementations, we performed the addition in a central accumulation unit by applying sequential processing of the input. The central accumulation unit 
increases the scalability and sequential processing enables preliminary carry canceling which improves the computation accuracy. Performance analysis revealed that the proposed design significantly reduces the hardware costs for high accuracy requirements and provides a high fault tolerance compared to a state-of-the-art design.

%\balance

%--------------------------------------------------------------------
% Acknowledgment (optional)
%--------------------------------------------------------------------
\ifdefined\ACK
  \section{Acknowledgment}
  \input{Acknowledgment}
\fi

% ===================================================================
% Appendices  
%\appendix \label{sec:Appendix}
%\section{Appendix 1}
%\input{Appendix}
%Appendix goes here \ldots
% ===================================================================

%--------------------------------------------------------------------
% References
%--------------------------------------------------------------------
%\nocite{*}
\bibliographystyle{IEEEtran}
\bibliography{IEEEabrv,References}

\end{document}